\documentclass[copyright,creativecommons]{eptcs}
\providecommand{\event}{LSFA 2012} 
\usepackage{breakurl}             
\usepackage{amsthm}
\usepackage{proof}
\usepackage{amsmath}
\usepackage{amssymb}

\title{A weak HOAS approach\\to the POPLmark Challenge}
\author{Alberto Ciaffaglione
\institute{Universit\`a di Udine, Italia\\
Dipartimento di Matematica e Informatica\\
via delle Scienze, 206 - 33100 Udine, Italia}
\email{alberto.ciaffaglione@uniud.it}
\and
Ivan Scagnetto
\institute{Universit\`a di Udine, Italia\\
Dipartimento di Matematica e Informatica\\
via delle Scienze, 206 - 33100 Udine, Italia}
\email{ivan.scagnetto@uniud.it}
}
\def\titlerunning{A weak HOAS approach to the POPLmark Challenge}
\def\authorrunning{A. Ciaffaglione \& I. Scagnetto}

\newcommand {\sF}{System F$_{<:}$}
\newcommand{\ccind}{\ensuremath{\text{CC}^\text{Ind}}}
\newcommand {\ie}{\textit{i.e}.}
\newcommand {\eg}{\textit{e.g}.}
\newcommand {\aka}{\textit{a.k.a}.}
\newcommand {\wrt}{\textit{w.r.t}.}

\newtheorem{definition}{Definition}
\newtheorem{theorem}{Theorem}
\newtheorem{lemma}{Lemma}
\newtheorem{proposition}{Proposition}

\begin{document}

\maketitle

\begin{abstract}
  Capitalizing on previous encodings and formal developments about
  nominal calculi and type systems, we propose a weak Higher-Order
  Abstract Syntax formalization of the type language of pure System
  F$_{<:}$ within Coq, a proof assistant based on the Calculus of
  Inductive Constructions.

  Our encoding allows us to accomplish the proof of the transitivity
  property of algorithmic subtyping, which is in fact the first of the
  three tasks stated by the POPLmark Challenge, a set of problems that
  capture the most critical issues in formalizing programming language
  metatheory.
\end{abstract}

\section{Introduction}
\label{sec:poplmark}

It is well known that formal proofs about programming language
metatheory and semantics are long and tedious, and that their
complexity is essentially due to the management of the details;
actually, it may happen that small mistakes or missed subtle cases
cause to invalidate large amounts of work, with this effect that
worsens as languages scale.  Automated proof assistants can help to
ease the problem, with several potential benefits: it may be simpler
to reuse work, to keep definitions and proofs consistent, to ensure a
firm relationship between theory and implementation.  Nevertheless, it
is apparent that computer-aided formal reasoning is not commonplace,
even for programming language designers and researchers.  Therefore,
the POPLmark Challenge \cite{poplmark} has proposed a framework and a
set of benchmarks for measuring the progress in the area, envising a
future in which research papers on programming languages will be
routinely accompanied by an electronic appendix with machine-checked
proofs.

The Challenge concerns a set of problems about the metatheory of a
variant of the System F$_{<:}$ \cite{DBLP:books/daglib/0005958}, \aka\
polymorphic, or second-order, lambda calculus. This choice has the
intent to pick out some features of programming languages that are
known to be difficult to formalize; in such a way, the problematic
aspects can be exploited to compare alternative technologies that have
been successfully experimented on specific areas.  In detail, the
Challenge concentrates on variable binding, complex recursion and
induction, definition and proof reuse, and experimentation of
generated sample programs.

In this paper we focus on the \emph{first task} among the three ones
in the Challenge suite, by considering \sF's \emph{type language}.
Essentially, such an object system features variable binding and
subtyping.
In fact, we adopt a methodology for encoding and reasoning formally on
\sF\ which takes most advantage of the features provided by logical
frameworks based on \emph{type theories}, and carry out our effort
within the Coq implementation \cite{coq} of the \emph{Calculus of
  Inductive Constructions} (\ccind)
\cite{coquandCC,DBLP:conf/tlca/Paulin-Mohring93}.

A common problem is that encoding and reasoning about a formal system
adds further complexity to already cumbersome judgments and proofs.
In order to be practically useful, therefore, it is important that the
formalization is as clean and compact as possible: ideally, most (if
not all) details implicitly taken for granted working with paper and
pencil should be automatically provided by the formal development.

A central point pursued by the Challenge is an efficient management of
inductively-defined structures with \emph{binders}. To this end, we
employ \emph{Higher-Order Abstract Syntax} (HOAS)
\cite{DBLP:conf/lics/HarperHP87,DBLP:conf/pldi/PfenningE88}, an
approach that uses binders \emph{in the metalanguage} to represent
binders in the object language, thus providing an high level of
abstraction.  More precisely, since we work in a type theory with
induction, we use \emph{weak} HOAS \cite{HMS-01}: binders are encoded
as \emph{second}-order term constructors that take as arguments
functions over a parametric, open (\ie, non-inductive) type of
\emph{variables}.  In this way, while we keep benefiting from
inductive definition and proof principles, we gain the advantage that
$\alpha$-conversion of abstractions and capture-avoiding substitution
of variables for variables are automatically ensured by the
metalanguage.

The main drawback of (weak) HOAS in \ccind\ (and Coq) is that it is
difficult to reason \emph{about} the encodings, because there is a
limited support for higher-order recursion and induction. To overcome
this problem, we adopt the \emph{Theory of Contexts} (ToC)
\cite{DBLP:conf/icalp/HonsellMS01}, a small set of axioms which can be
added to \ccind\ to represent some basic and natural properties of
variable and term \emph{contexts}.  It is apparent that we lose full
constructivity by using axioms; on the other hand, the ToC requires a
very low mathematical and logical overhead for porting to the formal
setting the arguments on paper.


In the end, by exploiting the above tools, we fulfill the first task
of the Challenge: that is, we prove the \emph{transitivity} and the
\emph{narrowing} properties of algorithmic subtyping for \sF.  We
believe that our result is relevant because the present one is the
\emph{first} weak HOAS approach to the Challenge\footnote{Among the
  proposed solutions collected by the Challenge web page
  \cite{poplmark:url}, our encoding is also the first HOAS one
  \emph{in Coq}.}, hence it provides extra feedback about the two
issues of representing and reasoning about binders and carrying out
formal proofs by mutual, nested structural induction on \sF's type
language.

In the next two sections, we recap the first task of the Challenge and
we rephrase it \emph{on paper} as a preliminary step towards its
formal treatment in \ccind.  In the core Section
\ref{sec:formalization} we present and discuss the formalization
itself (the Coq code is available at the web appendix of this paper
\cite{alberto:url}), and in the final sections we connect it to the
related literature and to the Challenge metrics of success.

\section{Algorithmic subtyping in the \sF}
\label{sec:systemF}

The POPLmark Challenge \cite{poplmark} addresses the metatheory of a
call-by-value variant of System F$_{<:}$, a calculus of moderate
scale. The first part of the Challenge, which we deal with in the
present paper, focuses just on the \emph{type language}, that we
consider in its \emph{pure} version here, \ie, without record types.

The syntax of \emph{types} features variables (taken, as usual, from
an infinite set of distinct symbols), the constant $Top$ (the
supertype of any type), functions, and bounded quantification (\ie,
universal types):
\[
\begin{array}{llcllll}
Type:\ & S,\ T & ::= & X & \quad \textrm{type variable}
                     & \qquad Top & \quad \textrm{maximal type}\\
      &        &     & S \to T & \quad \textrm{function type}
                     & \qquad \forall X{<:}S.\ T & \quad \textrm{universal type}
\end{array}
\]
Universal types, which are the individual characteristic of F$_{<:}$,
arise by combining polymorphism and subtyping: on the one hand types
such as $\forall X.\ T$ are intended to specify the type of
polymorphic functions; on the other hand bounded universal quantifiers
such as $\forall X{<:}S$ carry subtyping constraints.
Actually, the universal type $\forall X{<:}S.\ T$ has the effect of
binding the occurrence of $X$ in $T$, but not in $S$.

The \emph{type environments} are formed by subtyping constraints too,
involving type variables and types:
\[
\begin{array}{llcll}
Env:\ & \Gamma & ::= & \emptyset
                    & \qquad \textrm{empty type environment} \\
    &         &     & \Gamma,\ X{<:}T
                    & \qquad \textrm{type variable binding}\\
\end{array}
\]
Type variables within environments have to respect a \emph{scoping}
discipline: only fresh variables can be introduced, that is, $X
{\notin} dom(\Gamma)$; moreover, such variables cannot occur free in
the type they are bound to, \ie, $X {\notin} fv(T)$; finally, the
variables that appear free in $T$ have to be already collected in the
environment $\Gamma$.  Hence, a typical two-variable well-scoped
environment is $X{<:}Top,\ Y{<:}X$\footnote{We will give formal
 definitions of the mentioned concepts (well-known, though) in
 Section \ref{sec:sequents}.}.

\emph{Algorithmic} subtyping, $\Gamma \vdash S {<:} T$, captures the
intuition that ``$S$ is a subtype of $T$ under assumptions $\Gamma$'',
which means that ``an instance of $S$ may be safely used wherever an
instance of $T$ is expected''.
It is defined by induction and it is intended to concern only
\emph{well-scoped} types (\ie, when $\Gamma \vdash S {<:} T$ is
derived, all the type variables that occur free in $S$ and $T$ have to
be in the domain of $\Gamma$):
\[
\begin{array}{c}
\infer[(Top)]
{\Gamma \vdash S <: Top}
{}
\qquad
\infer[(Refl)]
{\Gamma \vdash X <: X}
{}
\qquad
\infer[(Trans)]
{\Gamma \vdash X <: T}
{X{<:}U \in \Gamma \quad \Gamma \vdash U <: T}
\\
\\
\infer[(Arr)]
{\Gamma \vdash S_1 \to S_2 <: T_1 \to T_2}
{\Gamma \vdash T_1 <: S_1 \quad \Gamma \vdash S_2 <: T_2}
\qquad
\infer[(All)]
{\Gamma \vdash \forall X{<:}S_1.\ S_2 <: \forall X{<:}T_1.\ T_2}
{\Gamma \vdash T_1 <: S_1 \quad \Gamma, X{<:}T_1 \vdash S_2 <: T_2}
\end{array}
\]

The Challenge focuses on the algorithmic version of subtyping because
its ultimate goal is the \emph{experimentation} of real
implementations of the formalized definitions. On the other hand,
being syntax-directed, algorithmic subtyping is easier \emph{to reason
with} than its equivalent, more familiar \emph{declarative}
presentation, where the rules $(Refl)$ and $(Trans)$ are replaced by
the following ones:
\[
\begin{array}{c}
\infer[(1)]
{\Gamma \vdash X <: U}
{X{<:}U \in \Gamma}
\qquad
\infer[(2)]
{\Gamma \vdash S <: S}
{}
\qquad
\infer[(3)]
{\Gamma \vdash S <: U}
{\Gamma \vdash S <: T \quad \Gamma \vdash T <: U}
\end{array}
\]

In fact, the first task of the Challenge addresses the relationship
between the two subtyping versions, as it consists to prove that the
\emph{transitivity} property $(3)$ is a derivable property within the
algorithmic system (the same holds for reflexivity $(2)$, which is not
problematic).

The proof of the transitivity is challenging essentially in two
respects: it has to be proved \emph{together} with the
\emph{narrowing} property, and such a proof requires a \emph{mutual
 and nested induction} proof argument.

\begin{proposition}[Transitivity and Narrowing]\label{challenge1A}

If $\Gamma \vdash S <: Q$ and $\Gamma \vdash Q <: T$, then $\Gamma
\vdash S <: T$.


If $\Gamma, X{<:}Q, \Delta \vdash M <: N$ and $\Gamma \vdash P <:
Q$, then $\Gamma, X{<:}P, \Delta \vdash M <: N$.
\begin{proof}
By induction on the structure of the type $Q$.

The proof for transitivity proceeds by an inner induction on the
structure of the derivation $\Gamma \vdash S <: Q$, with a case
analysis on the final rule of such a derivation and on that of the
second hypothesis $\Gamma \vdash Q <: T$.
We illustrate the crucial case when both the derivations end with an
application of the $(All)$ rule:
\[
 \infer[(All)]
 {\Gamma \vdash S \equiv \forall X{<:}S_1. S_2 <: \forall X{<:}Q_1. Q_2 \equiv Q}
 {\infer[]{\Gamma \vdash Q_1 <: S_1}{\vdots} \quad \infer[]{\Gamma, X{<:}Q_1 \vdash S_2 <: Q_2}{\vdots}}
\qquad
 \infer[(All)]
 {\Gamma \vdash Q \equiv \forall X{<:}Q_1. Q_2 <: \forall X{<:}T_1. T_2 \equiv T}
 {\infer[]{\Gamma \vdash T_1 <: Q_1}{\vdots} \quad \infer[]{\Gamma, X{<:}T_1 \vdash Q_2 <: T_2}{\vdots}}
\]
To conclude $\Gamma \vdash \forall X{<:}S_1. S_2 <: \forall X{<:}T_1.
T_2$ via the $(All)$ rule, two premises are needed: first, $\Gamma
\vdash T_1 <: S_1$ may be derived by induction hypothesis from the
third and the first subderivations; however, the induction hypothesis
cannot be applied to the second and fourth subderivations (to deduce
$\Gamma, X{<:}T_1 \vdash S_2 <: T_2$), because their environments are
different.
Hence, the narrowing property, \ie, the \emph{outer} induction
hypothesis (being $Q_1$ \emph{structurally smaller} than $Q$) has to
be exploited, to derive $\Gamma, X{<:}T_1 \vdash S_2 <: Q_2$ from the
second and the third subderivations.
Then, to construct the required derivation $\Gamma, X{<:}T_1 \vdash
S_2 <: T_2$ from this last hypothesis and the fourth subderivation, it
is necessary to apply again the \emph{outer} induction hypothesis (the
transitivity itself, with $Q_2$ structurally smaller than $Q$).

Similarly, the proof for narrowing proceeds by an inner induction on
the structure of the derivation $\Gamma, X{<:}Q, \Delta \vdash M <:
N$, again with a case analysis on the final rule applied.  The
treatment of this ``twin'' property is even subtler when the last rule
applied is $(Trans)$, and $M$ is exactly $X$:
\[
 \infer[(Trans)]
 {\Gamma, X{<:}Q, \Delta \vdash M \equiv X <: N}
 {\infer[]{\Gamma, X{<:}Q, \Delta \vdash Q <: N}{\vdots}}
\]
Now, $\Gamma, X{<:}P, \Delta \vdash Q <: N$ may be derived by
induction hypothesis, and $\Gamma, X{<:}P, \Delta \vdash P <: Q$ via a
straightforward \emph{weakening} property.
This time, the \emph{outer} induction hypothesis has to be exploited
\emph{with the same} $Q$; that is, the transitivity property is used
to deduce $\Gamma, X{<:}P, \Delta \vdash P <: N$ from the two inferred
derivations.
In the end, an application of the $(Trans)$ rule allows to obtain
$\Gamma, X{<:}P, \Delta \vdash X <: N$.

The present proof is reported in
\cite{poplmark,DBLP:books/daglib/0005958}, albeit not in a fully
detailed fashion.
\end{proof}
\end{proposition}

We notice finally that the presentation of \sF\ \cite{poplmark}, that
we have displayed and commented on, leaves implicit those aspects that
form the core of the Challenge: $\alpha$-conversion and
capture-avoiding substitution (as in standard practice), and the
well-scoping discipline (\emph{on purpose}).

\section{An alternative formulation of \sF}
\label{sec:sequents}

We give now an alternative presentation of \sF's subtyping, making
explicit some concepts that have been left implicit in the original
formulation reported in Section \ref{sec:systemF}. While carrying out
this step, we are mainly inspired by the features provided by logical
frameworks based on type theory.

We use here the same syntax for \emph{types} as in Section
\ref{sec:systemF}; on the other hand, we perform small changes on the
\emph{subtyping} system, and we prove that the new version is
equivalent to the original one.
The formalization in \ccind\ of the resulting system will be then
discussed in the following section.

We manage the \emph{type environment} as a concrete pair-component
collection, thus pursuing a \emph{sequent-style} encoding of
subtyping; consequently, we state formally two concepts related to the
environment itself.
First, we define the \emph{closure} of types $T$ \wrt\ environments
$\Gamma$ (a sort of compatibility) via the relation \emph{$closed
 \subseteq Type {\times} Env$}, to state that the free variables of
$T$ have to appear in the domain of $\Gamma$.
Further, the \emph{well-formedness} of environments $ok \subseteq Env$
prescribes that, when a new pair $\langle X, T \rangle$ makes an
environment $\Gamma$ grow, $X$ must both be fresh \wrt\ $\Gamma$ and
not appear in $T$, and $T$ has to be closed \wrt\ $\Gamma$.

In what follows, we write $fv(T)$ for the type variables occurring
free in a type $T$, and overload the symbols ``$\in,\notin$'' in a way
which is clear from the context.

\begin{definition}[Closure, Well-formedness]\label{closure}
 For $\Gamma {=} \langle X_1, T_1 \rangle, \ldots, \langle X_n, T_n
 \rangle $ an environment, $T$ a type, we define the \emph{domain} of
 $\Gamma$ and the predicates \emph{closed} and \emph{ok} as follows:
\[
\begin{array}{cc}
dom(\Gamma) \triangleq \{X_1, \ldots, X_n\}
& \qquad
closed(T,\Gamma) \triangleq \forall Y.\ Y {\in} fv(T) \Rightarrow \exists U.\ \langle Y, U \rangle {\in} \Gamma
\\
\\
\infer[(ok{\cdot}\emptyset)]
{ok(\emptyset)}
{}
& \qquad
\infer[(ok{\cdot}pair)]
{ok(\Gamma, \langle X, T \rangle)}
{ok(\Gamma) \quad X {\notin} dom(\Gamma) \quad closed(T,\Gamma)}
\end{array}
\]
\end{definition}
We notice that we do not need the condition $X {\notin} fv(T)$ among
the premises of the $(ok{\cdot}pair)$ rule, because it can be derived
from the second and the third hypotheses.
Finally, the main subtype judgment $\Gamma \vdash S {<:} T$ is
rendered as $sub(\Gamma, S, T)$, where $sub$ is a predicate defined on
$3$-tuples, $sub \subseteq Env {\times} Type {\times} Type$.

\begin{definition}[Subtyping]\label{new-subtyping}
 If $\Gamma$ is a type environment, $S,S_1,S_2,T,T_1,T_2,U$ types,
 then the predicate \emph{sub} is defined by induction, as follows:
\[
\begin{array}{c}
\infer[(top)]
{sub(\Gamma,\ S,\ Top)}
{ok(\Gamma) \qquad closed(S,\ \Gamma)}
\qquad
\infer[(var)]
{sub(\Gamma,\ X,\ X)}
{ok(\Gamma) \qquad \langle X, U \rangle \in \Gamma}
\\
\\
\infer[(trs)]
{sub(\Gamma,\ X,\ T)}
{\langle X, U \rangle \in \Gamma \qquad sub(\Gamma,\ U,\ T)}
\qquad
\infer[(arr)]
{sub(\Gamma,\ S_1 {\to} S_2,\ T_1 {\to} T_2)}
{sub(\Gamma,\ T_1,\ S_1) \qquad sub(\Gamma,\ S_2,\ T_2)}
\\
\\
\infer[(all)]
{sub(\Gamma,\ \forall X{<:}S_1. S_2,\ \forall X{<:}T_1. T_2)}
{sub(\Gamma,\ T_1,\ S_1) \qquad \textrm{for all } X,\ ok(\Gamma,\ \langle X, T_1 \rangle) \Rightarrow sub((\Gamma, \langle X, T_1 \rangle),\ S_2,\ T_2)}
\end{array}
\]
\end{definition}

It is apparent that our presentation of subtyping is equivalent to the
original one of Section \ref{sec:systemF}: informally arguing for such
an adequacy, we remark that we are using the same \emph{type
  environments} and that we have formalized their
\emph{well-formedness} and a kind of \emph{compatibility} between the
types and the environments themselves, two concepts which are implicit
in the POPLmark Challenge statement.

To prove \emph{formally} such an adequacy, we have to relate the
subtyping definitions in the two settings; this requires a preliminary
lemma, to connect each other the three judgments defined in this
section.
In the following, given an environment $\Gamma$, $perm(\Gamma)$ stands
for a permutation of its components.

\begin{lemma}[Auxiliary judgments]\label{pre-adequacy}
 For all $\Gamma {\in} Env$, and $S, T {\in} Type$:

1) $sub(\Gamma, S, T) \Rightarrow ok(\Gamma)$;

2) $sub(\Gamma, S, T) \Rightarrow closed(S,
\Gamma) \land closed(T, \Gamma)$.

\begin{proof} 1) By induction on the structure of the derivation of
  $sub(\Gamma, S, T)$.  2) By induction on the structure of the
  derivation of $sub(\Gamma, S, T)$, and point 1.
\end{proof}
\end{lemma}

\begin{theorem}[Adequacy]\label{adequacy}
 For all $\Gamma {\in} Env$, and $S, T {\in} Type$: $sub(\Gamma,S,T)$
 if and only if $\Gamma \vdash S <: T$.

\begin{proof}
 By structural induction on the hypothetical derivations, and Lemma
 \ref{pre-adequacy}.
\end{proof}
\end{theorem}

\begin{lemma}[Environment]\label{pre-challenge}
 For all $\Gamma, \Delta {\in} Env$, and $X, P, Q, S, T {\in} Type$:

 1) Well-formedness: $ok(\Gamma, \langle X, Q \rangle, \Delta) \land
 sub(\Gamma, P, Q) \Rightarrow ok(\Gamma, \langle X, P \rangle,
 \Delta)$;

 2) Permutation: $sub(\Gamma, S, T) \land ok(perm(\Gamma))
 \Rightarrow sub(perm(\Gamma), S, T)$;

 3) Weakening: $sub(\Gamma, S, T) \land ok(\Gamma, \Delta)
 \Rightarrow sub((\Gamma, \Delta), S, T)$.

 \begin{proof} 1) By induction on the structure of $\Delta$, and
   Lemma \ref{pre-adequacy}.2.  2) By induction on the derivation of
   $sub(\Gamma, S, T)$, and Lemma \ref{pre-adequacy}.1.  3) By
   induction on the derivation of $sub(\Gamma, S, T)$, and point 2.
\end{proof}
\end{lemma}

We are ready now to address the first Challenge, by ensuring that our
version of subtyping fulfills the reflexivity, transitivity and
narrowing properties.

\begin{proposition}[POPLmark Challenge, 1A]\label{challenge}
 For all $\Gamma, \Delta {\in} Env$, and $P, Q, M, N, S, T {\in}
 Type$:

\begin{description}
\item[Reflexivity] $ok(\Gamma) \land closed(S, \Gamma) \Rightarrow
 sub(\Gamma, S, S)$;

\item [Transitivity] $sub(\Gamma, S, Q) \land sub(\Gamma, Q, T)
 \Rightarrow sub(\Gamma, S, T)$;

\item [Narrowing] $sub((\Gamma, \langle X, Q \rangle, \Delta), M, N)
 \land sub(\Gamma, P, Q) \Rightarrow sub((\Gamma, \langle X, P
 \rangle, \Delta), M, N)$.
\end{description}

\begin{proof}
 \textbf{Reflexivity} By induction on the structure of $S$.

 As shown in Proposition \ref{challenge1A}, Transitivity and
 Narrowing are proved simultaneously by induction on the structure of
 $Q$; we point out here some extra details, which depend on the
 different cases of $Q$.

 \textbf{Transitivity} $Q {=} Top$: via Lemma~\ref{pre-adequacy}.2.
 $Q {=} Y$: by inner induction on the derivation of $sub(\Gamma, S,
 Y)$.  $Q {=} U {\to} V$: by inner induction on the derivation of
 $sub(\Gamma, S, U {\to} V)$, Lemma~\ref{pre-adequacy}.2, and the
 outer induction hypothesis, \ie, the transitivity statement itself
 twice, with $U$ and $V$, which are structurally smaller than $Q$.
 $Q {=} \forall Y{<:}U. V$: by inner induction on the derivation of
 $sub(\Gamma, S, \forall Y{<:}U. V)$, Lemma~\ref{pre-adequacy}.2, and
 the outer induction hypothesis, this time both the narrowing
 statement with $U$ and the transitivity with $V$, where, again, both
 $U$ and $V$ are structurally smaller than $Q$ (see also Proposition
 \ref{challenge1A}).

 \textbf{Narrowing} All the cases require an inner induction on the
 derivation of $sub((\Gamma, \langle X, Q \rangle, \Delta), M, N)$,
 and Lemmas \ref{pre-adequacy}.1, \ref{pre-challenge}.1. When the
 $(trs)$ rule is matched by such an inner induction, all the cases
 but the $Q {=} Top$ one need the application of the outer induction
 hypothesis, \ie, the transitivity statement with the starting $Q$
 (see also Proposition \ref{challenge1A}).
 Moreover, when $(trs)$ is matched, the $Q {=} Top$ case requires the
 Lemma \ref{pre-adequacy}.2, and the remaining cases the Weakening
 property (Lemma \ref{pre-challenge}.3).
\end{proof}
\end{proposition}

\section{Formalization of \sF\ in \ccind}
\label{sec:formalization}

When encoding a formal system in a type-theory based logical framework
(LF), one of the most tedious and time-consuming tasks is that of
representing variables and the related machinery of
$\alpha$-conversion and capture-avoiding substitution.  Traditional
solutions like, \eg, de~Bruijn indices and first-order variables,
force the user to spend a lot of time in formalizing and proving a
huge number of properties about free and bound occurrences of
variables, of $\alpha$-conversion, and involved concepts. Often, such
development greatly outweighs over the core part of the metatheory's
formalization.

An alternative approach, known as Higher-Order Abstract Syntax (HOAS)
\cite{DBLP:conf/lics/HarperHP87,DBLP:conf/pldi/PfenningE88}, has been
introduced for overcoming such an overhead. Its gist amounts to use
the metavariables of the LF to represent the variables of the object
language; in such a way, $\alpha$-conversion and capture-avoiding
substitution are completely delegated to the framework: in fact,
binders are modeled by functional constructors, and substitution is
modeled by functional application
\cite{DBLP:journals/jar/AvronHMP92,DBLP:conf/tlca/DespeyrouxFH95}.
Despite this apparent improvement, it is well known (see, \eg,
\cite{DBLP:conf/tlca/DespeyrouxFH95,mik:eltop}) that HOAS does not
cope well with inductive types, yielding several problems:
\begin{itemize}
\item Impossibility to adopt ``full'' HOAS representation of binders:
  functional types like $(T\rightarrow T)\rightarrow T$ violate the
  positivity constraints required by inductive constructors (thus, it
  is not possible to delegate the substitution of terms into terms to
  the metalevel).
\item Lack of suitable higher-order induction/recursion principles,
  which would allow to program with and reason about functional terms.
\item Impossibility to use inductive types, \eg, $Var$, to represent
  variables: otherwise, higher-order constructors (like
  $(Var\rightarrow T)\rightarrow T$) could generate ``exotic''
  parasite terms, \ie, terms not corresponding to any term of the
  object language.
\item Difficulty or impossibility to reason at the object level about
  the concepts and mechanisms delegated to the metalanguage.
\end{itemize}

Several attempts have been made to reconcile binding constructs with
induction principles, also via the design and implementation of new
logics (\eg, Nominal Logic
\cite{DBLP:journals/fac/GabbayP02,DBLP:journals/iandc/Pitts03} and
$FO\lambda^{\Delta\nabla}$ \cite{DBLP:journals/tocl/MillerT05}).
Although these solutions provide with advantages and support for a
suitable representation of variables and binders, they require the
user to switch to new and significantly different frameworks, to learn
them from scratch, and reimplement/translate the preceding work.

In this paper, we resort to a more ``conservative'' approach instead,
which has been already exploited in several case studies about the
encoding of process algebras, and static and dynamic semantics
\cite{HMS-01,DBLP:journals/entcs/HonsellMS01,DBLP:journals/entcs/ScagnettoM02,DBLP:conf/icfp/CiaffaglioneLM03,DBLP:conf/lpar/CiaffaglioneLM03,DBLP:journals/jar/CiaffaglioneLM07}.
Actually, we introduce in the Coq implementation \cite{coq} of \ccind\
type theory \cite{coquandCC,DBLP:conf/tlca/Paulin-Mohring93} a weak
HOAS formalization of \sF\ (Sections \ref{sec:synt-enc},
\ref{sec:subTp}) together with a compact axiomatization of simple
properties about variables, named the Theory of Contexts
(Section~\ref{sec:toc}).

\subsection{Encoding of syntax: types and type environments}
\label{sec:synt-enc}

In the following, \texttt{Var} is the \emph{non-inductive} type
representing \sF's \emph{(type) variables}; therefore we can represent
in Coq variables like $X$, $Y$, \ldots with metalanguage variables
\texttt{X}, \texttt{Y}, \ldots of type \texttt{Var}. Next, we define
the \emph{inductive} type \texttt{Tp} to represent \sF's \emph{types},
with four constructors for the maximal type, variables\footnote{Notice
  that \texttt{var} is declared as a coercion operator, which avoids
  to type explicitly the constructor, where a variable should stand
  for a term of type \texttt{Tp}.}, function and universal types
(compare with Section \ref{sec:systemF}):
{
\begin{verbatim}
Parameter Var: Set.
Inductive Tp: Set := var: Var -> Tp      | top: Tp  
                  |  arr: Tp -> Tp -> Tp | fa : Tp -> (Var -> Tp) -> Tp.
Coercion var: Var >-> Tp.
\end{verbatim}}
\noindent
This encoding, via a parametric type \texttt{Var} for variables and an
inductive type \texttt{Tp} for terms of the object system, is in fact
a \emph{weak} HOAS encoding. The constructor \texttt{fa}, which is
higher-order (as it takes as second argument a function from
\texttt{Var} to \texttt{Tp}), allows us to represent correctly \sF's
binder ``$\forall$'', by delegating to the Coq system the management
of the bound variable $X$ in the expression $\forall X{<:}S. T$. To be
more precise, if we denote with \texttt{S} the encoding of $S$ and
with \texttt{T[X]} the encoding of $T$ (where the occurrence of the
encoded bound variable \texttt{X}, corresponding to $X$, is explicitly
denoted by the square brackets), the representation of $\forall
X{<:}S. T$ is given by \texttt{(fa S (fun X:Var => T[X]))}.  Hence,
the variable \texttt{X} is bound by the metalanguage functional
construct \texttt{fun}; it follows that $\alpha$-conversion and
capture-avoiding substitution of variables for variables are
automatically dealt with by the metalanguage of Coq.

As remarked in Section \ref{sec:sequents}, in this paper we present an
``explicit'' encoding of \emph{type environments} $\Gamma$; these are
encoded as \emph{lists} of pairs, whose components belong to the types
\texttt{Var} and \texttt{Tp}, respectively: {
\begin{verbatim}
Definition envTp: Set := (list (Var * Tp)).
\end{verbatim}}
\noindent
This choice is quite intuitive and natural, except for the fact that
now, obviously, the environments grow ``toward the left'' (\ie, the
head of the list), while environments ``on paper'' grow toward the
right.


In order to reason about variables, types and type environments, we
need a set of auxiliary predicates that formalize the concepts defined
in Section \ref{sec:sequents}, \ie, the (non)occurrence of variables
into types, the freshness of variables/presence of pairs inside
environments, and the well-scoping of types \wrt\ the environments
themselves. First, we introduce the inductive predicates \texttt{isin}
and \texttt{notin}: 
{
\begin{verbatim}
Inductive isin (X:Var): Tp -> Prop := isin_var: isin X X
        | isin_arr: forall S T:Tp, isin X S \/ isin X T -> isin X (arr S T)
\end{verbatim}
\begin{verbatim}
        | isin_fa : forall S:Tp, forall U:Var->Tp,
          isin X S \/ (forall Y:Var, ~X=Y -> isin X (U Y)) -> isin X (fa S U).
Inductive notin (X:Var): Tp -> Prop := notin_top: notin X top
        | notin_var: forall Y:Var, ~X=Y -> notin X Y
        | notin_arr: forall S T:Tp, notin X S -> notin X T -> notin X (arr S T)
        | notin_fa : forall S:Tp, forall U:Var->Tp,
          notin X S -> (forall Y:Var, ~X=Y -> notin X (U Y)) -> notin X (fa S U).
\end{verbatim}}
\noindent
The intuitive meaning of \texttt{(isin X T)} is that the variable
\texttt{X} occurs free in \texttt{T}, $X {\in} fv(T)$ in Section
\ref{sec:sequents}, while \texttt{(notin X T)} stands for the opposite
concept, $X {\notin} fv(T)$.
The two definitions are syntax-driven, with just one introduction rule
for each constructor of type \texttt{Tp} (apart from the \texttt{top}
case for \texttt{isin}).

Concerning the environments, we formalize the freshness of a variable
$X {\notin} dom(\Gamma)$ (\texttt{Gfresh}), the presence of a
constraint $\langle X, T \rangle {\in} \Gamma$ (\texttt{isinG}), and
the closure of a type $closed(T,\Gamma)$ (\texttt{Gclosed}) \wrt\
them:
{
\begin{verbatim}
Inductive Gfresh (X:Var): envTp -> Prop := GfVoid: Gfresh X nil
        | GfGrow: forall G:envTp, forall Y:Var, forall T:Tp,
                  Gfresh X G -> ~X=Y -> Gfresh X (cons (Y,T) G).
Inductive isinG (X:Var) (T:Tp): envTp -> Prop :=
          checkG: forall G:envTp, forall y:Var, forall U:Tp,
                  (X=Y /\ T=U) \/ isinG X T G -> isinG X T (cons (Y,U) G).
Definition Gclosed (T:Tp) (G:envTp): Prop :=
           forall X:Var, (isin X T) -> exists U:Tp, isinG X U G.
\end{verbatim}}
\noindent
We can then state the inductive formulation of the well-formedness of
environments: 
{
\begin{verbatim}
Inductive okEnv: envTp -> Prop := okVoid: okEnv nil
        | okGrow: forall G:envTp, forall x:Var, forall T:Tp,
                  okEnv G -> Gfresh X G -> Gclosed T G -> okEnv (cons (X,T) G).
\end{verbatim}}

\subsection{Encoding of the subtyping relation}
\label{sec:subTp}

The representation of the subtyping relation, $sub$ in Section
\ref{sec:sequents}, follows closely its counterpart on the paper,
apart from the constructor for the universal type \texttt{sub\_fa},
which is accommodated via an hypothetical premise about a fresh,
locally quantified variable, which makes the encoding higher-order:
{
\begin{verbatim}
Inductive subTp: envTp -> Tp -> Tp -> Prop :=
          sub_top: forall G:envTp, forall S:Tp,
                   okEnv G -> Gclosed S G -> subTp G S top
        | sub_var: forall G:envTp, forall X:Var, forall U:Tp,
                   okEnv G -> isinG X U G -> subTp G X X
        | sub_trs: forall G:envTp, forall X:Var, forall U T:Tp,
                   isinG X U G -> subTp G U T -> subTp G X T
        | sub_arr: forall G:envTp, forall S1 S2 T1 T2:Tp,
                   subTp G T1 S1 -> subTp G S2 T2 ->
                   subTp G (arr S1 S2) (arr T1 T2)
        | sub_fa : forall G:envTp, forall S1 T1:Tp, forall S2 T2:Var->Tp,
                   subTp G T1 S1 ->
                   (forall X:Var, okEnv (cons (X,T1) G) ->
                                  subTp (cons (X,T1) G) (S2 X) (T2 X)) ->
                   subTp G (fa S1 S2) (fa T1 T2).
\end{verbatim}}
%

\subsection{The Theory of Contexts}
\label{sec:toc}

The \emph{Theory of Contexts} (ToC,
\cite{DBLP:conf/icalp/HonsellMS01,DBLP:journals/jfp/BucaloHMSH06}) is
a type-theoretic axiomatization which has been proposed to give a
metalogical account of the fundamental notions of \emph{variable} and
\emph{context}\footnote{Contexts are ``terms with holes'', where the
  holes can be filled in by variables.} as they appear in HOAS.
%
%
Moreover, when the ToC is instantiated in a weak HOAS setting, it is
compatible with the recursive and inductive environments provided by
type theory-based logical frameworks and their implementations.


In fact, the axioms of this theory aim to reflect in the logic some
fundamental and natural properties of object-level ``term contexts''
and ``variables'' (or ``names'', in some formal systems, like, \eg,
process algebras). The main advantages of this approach are that it
requires a very low mathematical and logical overhead, and that it can
be ``plugged'' in several existing proof environments without
requiring any redesign of such systems.
We present now the informal intended meaning of the ToC.

\begin{description}
\item[Decidability of equality over variables] For any variables $x$
  and $y$, it is always possible to decide whether $x=y$ or $x \neq y$
  (``$=$'' is Leibniz's equality).

\item[Freshness/Unsaturation] For any term $M$, there exists a
  variable $x$ which does not occur free in it (another interpretation
  is that there is no term containing/saturating all the variables).

\item[Extensionality] Two term contexts are equal if they are equal on
  a fresh variable; that is, if $M(x)=N(x)$ and $x \notin
  M(\cdot),N(\cdot)$, then $M=N$.

\item[$\beta$-expansion] It is always possible to split a term into a
  context applied to a variable; that is, given a term $M$ and a
  variable $x$, there exists a context $N(\cdot)$ such that $N(x)=M$
  and $x \notin N(\cdot)$.
\end{description}

The instantiation process is very simple and syntax-driven. First, we
state the following axiom (in fact the decidability is required for
each type representing variables, the sole \texttt{Var} in our case):
{
\begin{verbatim}
Axiom LEM_Var: forall X Y:Var, X=Y \/ ~X=Y.
\end{verbatim}}
\noindent
where the prefix \texttt{LEM} stands for \emph{Law of Excluded
  Middle}; indeed, this is the minimum classical flavour that we
require to reason about (free) occurrences of variables.  Such
assumption is very close to the common practice, when working on the
paper with nominal systems.

The formalization of the Freshness/Unsaturation for terms
of type \texttt{Tp} is straightforward too:
{
\begin{verbatim}
Axiom unsat: forall T:Tp, exists X:Var, notin X T.
\end{verbatim}}

Next we have the instantiations of extensionality (\texttt{tp\_ext})
and $\beta$-expansion (\texttt{tp\_exp}, \texttt{ho\_tp\_exp}).
Notice that we need the $\beta$-expansion both at the level of
first-order contexts (\ie, terms with one hole, \texttt{tp\_exp}) and at the level of
second-order contexts (terms with two holes,
\texttt{ho\_tp\_exp}):
{
\begin{verbatim}
Axiom tp_ext: forall X:Var, forall S T:Var->Tp,
              (notin_ho X S) -> (notin_ho X T) -> (S X)=(T X) -> S=T.
Axiom tp_exp: forall S:Tp, forall X:Var, 
              exists S': Var->Tp, (notin_ho X S') /\ S=(S' X).
Axiom ho_tp_exp: forall S:Var->Tp, forall X:Var, 
              exists S': Var->Var->Tp,
              (notin_ho X (fun Y:Var => (fa top (S' Y)))) /\ S=(S' X).
\end{verbatim}}
\noindent where \texttt{notin\_ho} is a simple definition built on top
of the predicate \texttt{notin}, stating that a variable does not
occur in a context:
{
\begin{verbatim}
Definition notin_ho:= fun X: Var => fun S: Var->Tp =>
                      forall Y: Var, ~X=Y -> (notin X (S Y)).
\end{verbatim}}

The properties formalized by the ToC have emerged from practical
reasoning about process algebras, and have been proved to be quite
useful in a number of situations\footnote{Their consistency has been
proved in~\cite{DBLP:journals/jfp/BucaloHMSH06}, starting from an
idea of M.~Hofmann \cite{Hofmann:1999:SAH:788021.788940}.}.
Ultimately, their combined effect is that of recovering the
capability of reasoning by structural induction over contexts. We
explain this fact by means of an individual example, about the
\emph{monotonicity} of the predicate \texttt{isin}, which is needed
for deriving the reflexivity of the subtyping relation (see
Section~\ref{sec:coq-popl}):
{
\begin{verbatim}
Lemma isin_mono: forall T:Var->Tp, forall X Y:Var, ~X=Y -> (isin X (T Y)) ->
                 (forall Z: Var, ~X=Z -> (isin X (T Z))).
\end{verbatim}}
\noindent
A direct way to prove the lemma would be by higher-order induction on
the structure of \texttt{T:Var->Tp}; however, Coq does not provide
such a principle.  Moreover, a na\"ive (\ie, first-order) induction on
\texttt{(T Y)} does not work, since there is no way to infer something
on the structure of the context \texttt{T} from the structure of
\texttt{(T Y)} (notice that \texttt{Y} can occur free in \texttt{T}).
Hence, we prove a preliminary lemma: 
{
\begin{verbatim}
Lemma pre_isin_mono: forall n:nat, forall T:Tp, (lntp T n) ->
      forall Z:Var, forall U:Var->Tp, (notin_ho Z U) -> T=(U Z) ->
      forall X Y:Var, ~X=Y -> (isin X (U Y)) ->
      forall V:Var, ~X=V -> (isin X (U V)).
\end{verbatim}}
\noindent
where \texttt{lntp} is the predicate which counts the number of
constructors involved in a term of type \texttt{Tp}:
{
\begin{verbatim}
Inductive lntp: Tp -> nat -> Prop := 
     lntp_top : (lntp top (S O))
   | lntp_var : forall X:Var, (lntp X (S O))
   | lntp_arr : forall T T':Tp, forall n1 n2:nat,
                (lntp T n1) -> (lntp T' n2) -> 
                (lntp (arr T T') (S (plus n1 n2)))
   | lntp_fa  : forall T:Tp, forall U:Var->Tp, forall n1 n2:nat, 
                (lntp T n1) -> (forall X:Var, (lntp (U X) n2)) ->
                (lntp (fa T U) (S (plus n1 n2))).
\end{verbatim}}
\noindent
Therefore, \texttt{(lntp T n)} states that the term \texttt{T} is
``built'' using \texttt{n} constructors of the inductive type
\texttt{Tp}. This fact allows us to argue by \emph{complete} induction
on \texttt{n} in the proof of \texttt{pre\_isin\_mono}, thus
recovering the structural information about \texttt{T} via inversion
of the instance \texttt{(lntp T n)}. So far, we can apply
$\beta$-expansion to infer the existence of a context
\texttt{T':Var->Tp} such that \texttt{T=(T' z)}, where \texttt{z} does
not occur free in \texttt{T'}. Then, by applying the extensionality
property, we can deduce that \texttt{U=T'} and, since \texttt{T'} is
not a variable but a concrete $\lambda$-abstraction, we ``lift''
structural information to the level of functional terms. Such an
information can be finally used to solve the current goal,
\texttt{isin\_mono} in the case.

In order to be more concrete, let us consider the case where
\texttt{(lntp (T z) 1)} holds. By inverting such an hypothesis, we get
the case (among other ones) where the equality \texttt{(T z)=top}
holds. Then, we apply $\beta$-expansion (\texttt{tp\_exp}) to
\texttt{top}, yielding a context \texttt{T'=(fun x:Tp => top)}; in
particular, we can state that \texttt{(T z)=top=(T' z)}, whence we
infer \texttt{(T z)=((fun x:Tp => top) z)}. Finally, by means of the
extensionality axiom (\texttt{tp\_ext}), we ``lift'' such structural
information to higher-order terms: namely, we deduce \texttt{T=(fun
  x:Tp => top)}, \ie, we get the structural information we need about
\texttt{T}.


\subsection{Formal development of the POPLmark Challenge}
\label{sec:coq-popl}

In this section we illustrate the formal development carried out in
the Coq system in order to solve the first task of the POPLmark
Challenge, \ie\ reflexivity, transitivity (and narrowing) of subtyping
(Proposition \ref{challenge}). We start by introducing some auxiliary
lemmas; the mostly used property is the following: {
\begin{verbatim}
Lemma Gclosed_lemma: forall G:envTp, forall S T:Tp,
                     subTp G S T -> Gclosed S G /\ Gclosed T G.
\end{verbatim}}
\noindent
The informal meaning is that, if we derive \texttt{(subTp G S T)}
(under such an hypothesis we are able to deduce that \texttt{G} is a
well-formed environment, Lemma \ref{pre-adequacy}.1 of Section
\ref{sec:sequents}), then all the variables occurring free in
\texttt{S} and \texttt{T} belong to the domain of \texttt{G}. The
proof is carried out by induction on the derivation of \texttt{(subTp
  G S T)}, using \texttt{unsatG} when we need a variable which is
fresh \wrt\ the environment \texttt{G}:
{
\begin{verbatim}
Lemma unsatG: forall G:envTp, exists X:Var, Gfresh X G.
\end{verbatim}}
\noindent
As the reader may guess, the proof of \texttt{unsatG} relies heavily
upon the axiom \texttt{unsat} of the Theory of Contexts (see Section
\ref{sec:toc}). Actually, given an environment \texttt{G}, the idea is
just to scan the variable declaration list \texttt{(X1,T1)}, \ldots,
\texttt{(Xn,Tn)} in \texttt{G}, to build an arrow type \texttt{(arr X1
  (arr ...  (arr Xn top) ... ))}.  Then, by eliminating \texttt{unsat}
on this type, we can get a fresh variable not occurring into such type
and, consequently, not appearing in the domain of \texttt{G}:
{
\begin{verbatim}
Lemma domGtoT_notin: forall G:envTp, forall X:Var,
                     notin X (domGtoT G) -> Gfresh X G.
\end{verbatim}}
\noindent
where \texttt{domGtoT} is a function, defined by recursion on the
environment \texttt{G}, which builds the mentioned arrow type from the
variables belonging to its domain:
{
\begin{verbatim}
Fixpoint domGtoT (G:envTp):= match G with 
         | nil => top | (X,T)::G' => (arr X (domGtoT G')) end.
\end{verbatim}}
\noindent
The proof of \texttt{domGtoT\_notin} is performed by induction on the
structure of \texttt{G}, using the axiom \texttt{LEM\_Var} to
discriminate between the occurrences of variables.

Coming in the end to the POPLmark Challenge properties, the
\emph{reflexivity} requires that the type environment is well-formed
and the type under investigation is closed \wrt\ the environment
itself:
{
\begin{verbatim}
Lemma reflexivity: forall T:Tp, forall G:envTp, 
                   okEnv G -> Gclosed T G -> subTp G T T.
\end{verbatim}}
\noindent
The proof is a straightforward induction on the structure of
\texttt{T}, resorting to \texttt{LEM\_Var} when it is needed to
discriminate between free variables, and using the monotonicity of the
``occurrence'' predicate \texttt{isin}.

\emph{Transitivity} and \emph{narrowing} are proved together (as on
the paper), via an outer induction on the structure of the type
\texttt{Q}, which is then isolated in front of the two properties:
{
\begin{verbatim}
Theorem trans_narrow: forall Q:Tp,
  (forall S:Tp, forall G:envTp, 
  (subTp G S Q) -> forall T:Tp, (subTp G Q T) -> (subTp G S T))
  /\
  (forall G':envTp, forall M N:Tp, 
  (subTp G' M N) -> forall D G:envTp, forall X:Var, forall P:Tp, 
                    G'=(app D (cons (X,Q) G)) -> subTp G P Q ->
                    subTp (app D (cons (X,P) G)) M N).
\end{verbatim}}
\noindent
The proof of transitivity is, apart from the use of the Theory of
Contexts, similar to that on the paper, via an inner induction on the
derivation of \texttt{(subTp G S Q)}.

The same remark holds about the narrowing, whose management needs an
inner induction on the derivation of \texttt{(subTp G' M N)}, where
the environment \texttt{G'} is Coq's list \texttt{(app D (cons (X,Q)
  G))}, which is built by means of the \emph{append} function
\texttt{app}.
However, the narrowing requires two extra efforts.

First, as its formulation involves a structured environment, it has
been necessary to prove a series of technical lemmas involving Coq's
\emph{lists} and their relationship with the predicates
\texttt{Gfresh}, \texttt{isinG}, \texttt{Gclosed}, \texttt{okEnv}.
In carrying out such proofs, we have taken partial advantage of Coq's
built-in list library, especially about \emph{permutations}, which are
required by the Weakening property (Lemma \ref{pre-challenge}.3).

To master the sophisticated interdependence between the outer and the
inner structural inductions within the narrowing proof, we have
exploited a slight elaboration of ``modus ponens'': $\forall A,B:
Prop.\ A \land (A \Rightarrow B) \Rightarrow A \land B$ (where $A$ and
$B$ are intended to play the role of transitivity and narrowing,
respectively).
In fact, when the inner induction hypothesis for narrowing matches the
rule \texttt{sub\_trs} (see the proof of Proposition \ref{challenge}),
the outer induction hypothesis (\ie, transitivity) has to be applied
with the \emph{starting} \texttt{Q}, not with a structurally smaller
type.
Therefore, to handle the involved cases within the outer induction
(all but the \texttt{Q=top} one), we reduce to prove the transitivity
alone and the narrowing with the proof context enriched by the
transitivity additional hypothesis, instead of merely splitting the
two main proofs.



\section{Related work}
\label{sec:related}

At the time of writing, the POPLmark web page \cite{poplmark:url}
collects fifteen contributions, included ours.
In this section we give a brief account of the different approaches,
filtering them through the perspective of the \emph{first task} of the
Challenge.
Notice that we do not discuss here those works that employ the
\emph{pure} de Bruijn representation, because, according to the
POPLmark document \cite{poplmark}, it violates the ``reasonable
overhead'' primary metric of success test.
Nevertheless, de Bruijn's technique can be taken into account to
measure the progress of alternative representations, and its positive
sides may be combined to novel ones.

An approach that keeps de Bruijn indices to represent bound variables,
together with (first-order) names to manage free variables, is known
as \emph{locally nameless} representation.
This was first experimented in Coq by Leroy
\cite{leroy:coq,Leroy-POPLmark}, then refined by Chlipala
\cite{chlipala:coq}, Chargu\'eraud
\cite{chargueraud:coq,chargueraud-11-ln}, and ported to the
\texttt{Matita} proof assistant by Ricciotti \cite{ricciotti:matita}.
As de Brujin indices represent variables by positions relative to the
enclosing binders, there is no need to introduce $\alpha$-equivalence
for bound variables; on the other hand, two substitutions of types
(for indices and names) have to be managed. Explicit environments are
defined, and well-formedness of environments and types are introduced
to describe the main subtyping concept.

The opposite encoding choice is made by Stump \cite{stump:coq}, who
represents in Coq bound variables via names and free variables via de
Bruijn indices, by taking advantage from the Barendregt variable
convention, which assumes that bound and free variables come from
\emph{disjoint} sets.

\emph{Higher-Order Abstract Syntax (HOAS)} encodings are closer to
ours; we find an hybrid solution in ATS (commented on later in the
section), and two full HOAS formalizations, in Abella and Twelf.

The work carried out by Gacek in Abella \cite{gacek:abella,gacek:phd}
introduces a canonical HOAS representation of \sF's types (notice, in
particular, the signature of the universal constructor ``$\forall$'',
named \texttt{all}):
\begin{verbatim}
ty    type.
top   ty.
arrow ty -> ty -> ty.
all   ty -> (ty -> ty) -> ty.
\end{verbatim}
Since variables are represented by metavariables of type \texttt{ty},
the extra specification logic judgment \texttt{bound:ty->ty->o} has to
be defined to cope with the environment assumptions, and a
(simplified) environment well-formedness predicate
\texttt{ctx:olist->prop} is introduced to reason about subtyping.
Finally, to make structural induction on \sF's types feasible, a
predicate \texttt{wfty:ty->prop} is added.

The formalization carried out at Carnegie Mellon University within the
Twelf system \cite{cmu:twelf} uses the same signature for the syntax
of \sF's types (here, the universal constructor is named
\texttt{forall}):
\begin{verbatim}
tp: type. ...
forall: tp -> (tp -> tp) -> tp.
\end{verbatim}
Again, the environment assumptions require a distinguished judgment,
\texttt{assm:tp->tp->type}, but, differently from the Abella approach,
there is no explicit environment to reason on subtyping; instead, an
extra judgment \texttt{var:tp->type} is defined, to ``mark'' the types
which play the role of variables.

Summing up, variables are represented by Abella's and Twelf's
metavariables belonging to the types \texttt{ty} and \texttt{tp},
which are introduced to encode the syntax of System F$_{<:}$'s types.
Differently, we adopt a weak HOAS approach, by choosing a separate,
parametric type \texttt{Var} for representing variables:
\begin{verbatim}
Parameter Var: Set.
Inductive Tp: Set := ...
fa: Tp -> (Var -> Tp) -> Tp.
\end{verbatim}
In this way, we keep the advantage of delegating $\alpha$-conversion
and substitution of variables for variables to the metalanguage, while
retaining Coq's built-in induction principle for \texttt{Tp}. Of
course, in Abella and Twelf one has the extra possibility of
delegating the substitution of \emph{types} for variables, while we
should write an ad-hoc predicate.  However, this kind of substitution
is not required to deal with subtyping.


Also the solution proposed by Urban and coworkers in Isabelle
\cite{urban:isabelle}, and based on the \emph{Nominal (Logic)}
datatype package, is quite related to our approach. The signature of
types is the following:

\vspace{3mm}
\noindent
$
\begin{array}{l}
\mathbf{atom-decl}\ tyvrs\\
\mathbf{nominal-datatype}\ ty\ =\\
\ \ Tvar\ tyvrs\\
|\ Top\\
|\ Arrow\ ty\ ty\qquad (-\ \rightarrow\ -\ [100,\ 100]\ 100)\\
|\ Forall\ \ll tyvrs\gg\ ty\ ty
\end{array}
$

\vspace{3mm}
\noindent In this formalization type variables are represented by
atoms, therefore \sF's ``$\forall$'' binder is encoded via the
abstraction operator $\ll\ldots\gg$\ldots; this allows to prove that
$\alpha$-equivalent types are equal.  Then, a measure on the size of
types and the notion of capture-avoiding substitution are defined.

We remark that the intrinsic concepts of \emph{finite support} and
\emph{freshness} play in Nominal Logic a role which is similar to that
of occurrence (\texttt{isin}) and non-occurrence (\texttt{notin})
predicates, which are bundled with our axioms of the Theory of
Contexts (ToC). Actually, this is not fortuitous, since
in~\cite{DBLP:conf/icfp/MiculanSH05} the relation between the
intuitionistic Nominal Logic and the Theory of Contexts is clearly
explained by means of a translation of terms, formulas and judgments
of the former into terms and propositions of the \ccind, via a weak
HOAS encoding. It turns out that the (translation of the) axioms and
rules of the intuitionistic Nominal Logic are derivable in \ccind\
extended with the Theory of Contexts (\ccind\ + ToC).

An alternative high-level encoding technique exploits nested datatypes
for representing the variable binder in Coq
\cite{hirsch-mag:coq,DBLP:journals/jar/HirschowitzM12}. This approach,
whose characteristic feature is the encoding of the ``$\forall$''
operator (\texttt{Uni} in the following predicate), is named
\emph{nested abstract syntax} by its authors:
\begin{verbatim}
Inductive ftype (V:Type): Type := ...
| Uni: ftype V -> ftype ^V -> ftype V.
\end{verbatim}
where the type \verb|^|\texttt{V}, rendered by the \texttt{option}
datatype, denotes \texttt{V} extended with a new ``fresh'' element:
\begin{verbatim}
Inductive option (V:Type): Type := Some: V -> option V
                                |  None: option V.
\end{verbatim}
The main advantages consist of retaining the induction and recursion
principles provided by Coq and providing a categorical interpretation
of the whole approach. On the other hand, as the heavy use of
dependent typing is not always supported by Coq, ad-hoc techniques
have to be picked out.

Xi's \emph{hybrid} solution in ATS \cite{xi:ats} combines HOAS (for
types) with de~Bruijn indices (for environments).  ATS is a powerful
programming language, featuring dependent and linear types and
supporting theorem proving. However, in this case it is not possible
to state a meaningful comparison with our work, because Xi's
contribution does not address the first part of the Challenge.


Another approach which addresses a different part of the Challenge is
due to Fairbairn and carried out in
Alpha-Prolog~\cite{fairbairn:alpha}, with a complementary parser and
pretty printer written in OCaml. More precisely, this work provides a
nominal-style formalization of \sF\ with records and patterns,
allowing the user to ``animate'' the language, \ie, to explore the
language properties on specific examples.

\section{Conclusion}
\label{sec:conclusion}

Carrying out our weak HOAS formalization of System F$_{<:}$'s pure
type language in Coq, we have tried to stick to the POPLmark primary
metrics of success (see~\cite{poplmark}).
\begin{itemize}
\item \textit{Correctness.} In Section~\ref{sec:sequents} we have
  given an alternative presentation of \sF's subtyping concept, thus
  yielding a system which is equivalent to the original one (as stated
  by Theorem~\ref{adequacy}), but at the same time closer to the final
  formalization in \ccind.  In other words, the translation from the
  system ``on paper'', presented in Section~\ref{sec:sequents}, to its
  formal counterpart in Section~\ref{sec:formalization} is, except for
  the use of weak HOAS, a matter of syntactic sugar.

\item \textit{Reasonable overhead.} The weak HOAS encoding approach,
  together with the (suitable instantiation of the) Theory of
  Contexts, provides a smooth treatment of the (type) variable binder,
  and frees the user from the burden of dealing with low-level
  mechanisms about variables\footnote{Namely, $\alpha$-conversion and
    capture-avoiding substitution of variables for variables.}. In
  fact, bound variables are automatically dealt with by the
  metalanguage of Coq, which transparently renames them to avoid
  clashes with free ones. At the same time, our formalization allows
  the user to keep benefiting from the inductive features of \ccind,
  that is, recursion and induction principles.  Remarkably, the Theory
  of Contexts grants the extra ability to handle and reason about
  contexts (\ie, higher-order terms), lifting structural information
  to the level of functional terms.

\item \textit{Transparent technology.} In our opinion, both the formal
  representation of \sF's type language and the encoding of
  fundamental theorems' statements are easily readable and very close
  to their informal counterparts. Even the axioms of the Theory of
  Contexts are reminiscent of properties that are commonly taken for
  granted, working with ``paper and pencil''.

\item \textit{Reasonable cost of entry.} The Coq system is one of the
  most used proof assistants based on type theory; it is
  well-documented, and the provided tutorial allows everyone who is
  knowledgeable about programming language theory to use fruitfully
  the proof assistant, after a reasonable training effort, for the
  goals within the Challenge. More specifically, the Theory of
  Contexts may be injected in Coq without the need of any redesign of
  the system; moreover, as we have already pointed out, such a theory
  is rather easy to add on top of a signature, since it is
  syntax-driven.
\end{itemize}

Concluding, we stress that, even we have not pursued neither
optimization (of our encoding) nor competition (with the alternative
ones)\footnote{In fact, our contribute is the \emph{first} weak HOAS
  solution submitted to the POPLmark Challenge: as such, the spirit of
  our work is essentially to close a gap, and at the same time a first
  effort towards more ambitious goals (stated by the Challenge).}, our
formalization is still effective and very terse, in spite of lack of
support for HOAS encodings in Coq.  Actually, the source code of the
development preliminary to the main goal is 33.4~KB long, including
12.7~KB required to manage the type environment; also the main proof
(Reflexivity, Transitivity and Narrowing) is rather compact (it is
about 16~KB long), and it follows closely the trace of its
``informal'' counterpart, carried out ``on paper''.

From a pragmatic point of view, we want just to add two remarks.

First, we have suffered a little from the lack of ``smart'' support
for nested inductions, having to rearrange the goal statement and to
enrich it with suitable equalities, to correctly ``purge'' the
inconsistent cases automatically generated by the nested application
of the \texttt{induction} tactic.

Second, we have spent almost the $40\%$ of the preliminary script to
handle the type environment, which could be seen as an overhead. In
fact, we plan to investigate in future work the possibility to drop
the list machinery used to represent the type environment, by adopting
instead the \emph{bookkeeping technique}
\cite{mik:eltop,DBLP:conf/icfp/CiaffaglioneLM03,DBLP:conf/lpar/CiaffaglioneLM03,DBLP:journals/jar/CiaffaglioneLM07},
with a ``global'' environment and local hypotheses modeled via
hypothetic judgments.

\bibliographystyle{eptcs}

\bibliography{biblio}

\begin{thebibliography}{10}
\providecommand{\bibitemdeclare}[2]{}
\providecommand{\surnamestart}{}
\providecommand{\surnameend}{}
\providecommand{\urlprefix}{Available at }
\providecommand{\url}[1]{\texttt{#1}}
\providecommand{\href}[2]{\texttt{#2}}
\providecommand{\urlalt}[2]{\href{#1}{#2}}
\providecommand{\doi}[1]{doi:\urlalt{http://dx.doi.org/#1}{#1}}
\providecommand{\bibinfo}[2]{#2}

\bibitemdeclare{misc}{cmu:twelf}
\bibitem{cmu:twelf}
\bibinfo{author}{M. \surnamestart Ashley-Rollman\surnameend},
  \bibinfo{author}{K. \surnamestart Crary\surnameend} \&
  \bibinfo{author}{R. \surnamestart Harper\surnameend}:
  \emph{\bibinfo{title}{A solution to the POPLmark Challenge}}.
\newblock Available at [4].

\bibitemdeclare{article}{DBLP:journals/jar/AvronHMP92}
\bibitem{DBLP:journals/jar/AvronHMP92}
\bibinfo{author}{A. \surnamestart Avron\surnameend}, \bibinfo{author}{F.
  \surnamestart Honsell\surnameend}, \bibinfo{author}{I.~A. \surnamestart
  Mason\surnameend} \& \bibinfo{author}{R. \surnamestart
  Pollack\surnameend} (\bibinfo{year}{1992}): \emph{\bibinfo{title}{Using Typed
  Lambda Calculus to Implement Formal Systems on a Machine}}.
\newblock {\sl \bibinfo{journal}{J. Autom. Reasoning}}
  \bibinfo{volume}{9}(\bibinfo{number}{3}), pp. \bibinfo{pages}{309--354}.
\newblock \urlprefix\url{http://dx.doi.org/10.1007/BF00245294}.

\bibitemdeclare{inproceedings}{poplmark}
\bibitem{poplmark}
\bibinfo{author}{B.~E. \surnamestart Aydemir\surnameend\ et al.},
  (\bibinfo{year}{2005}): \emph{\bibinfo{title}{Mechanized Metatheory for the
  Masses: The PoplMark Challenge}}.
\newblock In \bibinfo{editor}{Joe \surnamestart Hurd\surnameend} \&
  \bibinfo{editor}{Thomas~F. \surnamestart Melham\surnameend}, editors: {\sl
  \bibinfo{booktitle}{TPHOLs}}, {\sl \bibinfo{series}{Lecture Notes in Computer
  Science}} \bibinfo{volume}{3603}, \bibinfo{publisher}{Springer}, pp.
  \bibinfo{pages}{50--65}.
\newblock \urlprefix\url{http://dx.doi.org/10.1007/11541868_4}.

\bibitemdeclare{misc}{poplmark:url}
\bibitem{poplmark:url}
\bibinfo{author}{B.~E. \surnamestart Aydemir\surnameend\ et al.},
  \emph{\bibinfo{title}{The POPLmark Challenge}}.
\newblock \urlprefix\url{http://www.seas.upenn.edu/~plclub/poplmark/}.

\bibitemdeclare{article}{DBLP:journals/jfp/BucaloHMSH06}
\bibitem{DBLP:journals/jfp/BucaloHMSH06}
\bibinfo{author}{A. \surnamestart Bucalo\surnameend}, \bibinfo{author}{F.
  \surnamestart Honsell\surnameend}, \bibinfo{author}{M. \surnamestart
  Miculan\surnameend}, \bibinfo{author}{I. \surnamestart
  Scagnetto\surnameend} \& \bibinfo{author}{M. \surnamestart
  Hofmann\surnameend} (\bibinfo{year}{2006}): \emph{\bibinfo{title}{Consistency
  of the theory of contexts}}.
\newblock {\sl \bibinfo{journal}{J. Funct. Program.}}
  \bibinfo{volume}{16}(\bibinfo{number}{3}), pp. \bibinfo{pages}{327--372}.
\newblock \urlprefix\url{http://dx.doi.org/10.1017/S0956796806005892}.

\bibitemdeclare{misc}{chargueraud:coq}
\bibitem{chargueraud:coq}
\bibinfo{author}{A. \surnamestart Chargu{\'e}raud\surnameend}:
  \emph{\bibinfo{title}{A solution to the POPLmark Challenge}}.
\newblock Available at [4].

\bibitemdeclare{article}{chargueraud-11-ln}
\bibitem{chargueraud-11-ln}
\bibinfo{author}{A. \surnamestart Chargu{\'e}raud\surnameend}
  (\bibinfo{year}{2011}): \emph{\bibinfo{title}{The Locally Nameless
  Representation}}.
\newblock {\sl \bibinfo{journal}{Journal of Automated Reasoning}}, pp.
  \bibinfo{pages}{1--46}.
\newblock \urlprefix\url{http://dx.doi.org/10.1007/s10817-011-9225-2}.

\bibitemdeclare{misc}{chlipala:coq}
\bibitem{chlipala:coq}
\bibinfo{author}{A. \surnamestart Chlipala\surnameend}:
  \emph{\bibinfo{title}{A solution to the POPLmark Challenge}}.
\newblock Available at [4].

\bibitemdeclare{inproceedings}{DBLP:conf/lpar/CiaffaglioneLM03}
\bibitem{DBLP:conf/lpar/CiaffaglioneLM03}
\bibinfo{author}{A. \surnamestart Ciaffaglione\surnameend},
  \bibinfo{author}{L. \surnamestart Liquori\surnameend} \&
  \bibinfo{author}{M. \surnamestart Miculan\surnameend}
  (\bibinfo{year}{2003}): \emph{\bibinfo{title}{Imperative Object-Based Calculi
  in Co-inductive Type Theories}}.
\newblock In \bibinfo{editor}{Moshe~Y. \surnamestart Vardi\surnameend} \&
  \bibinfo{editor}{Andrei \surnamestart Voronkov\surnameend}, editors: {\sl
  \bibinfo{booktitle}{LPAR}}, {\sl \bibinfo{series}{Lecture Notes in Computer
  Science}} \bibinfo{volume}{2850}, \bibinfo{publisher}{Springer}, pp.
  \bibinfo{pages}{59--77}.
\newblock \urlprefix\url{http://dx.doi.org/10.1007/978-3-540-39813-4_4}.

\bibitemdeclare{inproceedings}{DBLP:conf/icfp/CiaffaglioneLM03}
\bibitem{DBLP:conf/icfp/CiaffaglioneLM03}
\bibinfo{author}{A. \surnamestart Ciaffaglione\surnameend},
  \bibinfo{author}{L. \surnamestart Liquori\surnameend} \&
  \bibinfo{author}{M. \surnamestart Miculan\surnameend}
  (\bibinfo{year}{2003}): \emph{\bibinfo{title}{Reasoning on an imperative
  object-based calculus in Higher Order Abstract Syntax}}.
\newblock In: {\sl \bibinfo{booktitle}{MERLIN}}, \bibinfo{publisher}{ACM}, pp.
  \bibinfo{pages}{1--10}.
\newblock \urlprefix\url{http://doi.acm.org/10.1145/976571.976574}.

\bibitemdeclare{article}{DBLP:journals/jar/CiaffaglioneLM07}
\bibitem{DBLP:journals/jar/CiaffaglioneLM07}
\bibinfo{author}{A. \surnamestart Ciaffaglione\surnameend},
  \bibinfo{author}{L. \surnamestart Liquori\surnameend} \&
  \bibinfo{author}{M. \surnamestart Miculan\surnameend}
  (\bibinfo{year}{2007}): \emph{\bibinfo{title}{Reasoning about Object-based
  Calculi in (Co)Inductive Type Theory and the Theory of Contexts}}.
\newblock {\sl \bibinfo{journal}{J. Autom. Reasoning}}
  \bibinfo{volume}{39}(\bibinfo{number}{1}), pp. \bibinfo{pages}{1--47}.
\newblock \urlprefix\url{http://dx.doi.org/10.1007/s10817-006-9061-y}.

\bibitemdeclare{misc}{alberto:url}
\bibitem{alberto:url}
\bibinfo{author}{A. \surnamestart Ciaffaglione\surnameend} \&
  \bibinfo{author}{I. \surnamestart Scagnetto\surnameend}
  (\bibinfo{year}{2012}): \emph{\bibinfo{title}{A solution to the POPLmark Challenge}}.
\newblock Available at [4].

\bibitemdeclare{article}{coquandCC}
\bibitem{coquandCC}
\bibinfo{author}{T. \surnamestart Coquand\surnameend} \&
  \bibinfo{author}{G.~P. \surnamestart Huet\surnameend}
  (\bibinfo{year}{1988}): \emph{\bibinfo{title}{The Calculus of
  Constructions}}.
\newblock {\sl \bibinfo{journal}{Inf. Comput.}}
  \bibinfo{volume}{76}(\bibinfo{number}{2/3}), pp. \bibinfo{pages}{95--120}.
\newblock \urlprefix\url{http://dx.doi.org/10.1016/0890-5401(88)90005-3}.

\bibitemdeclare{inproceedings}{DBLP:conf/tlca/DespeyrouxFH95}
\bibitem{DBLP:conf/tlca/DespeyrouxFH95}
\bibinfo{author}{J. \surnamestart Despeyroux\surnameend},
  \bibinfo{author}{A.~P. \surnamestart Felty\surnameend} \&
  \bibinfo{author}{A. \surnamestart Hirschowitz\surnameend}
  (\bibinfo{year}{1995}): \emph{\bibinfo{title}{Higher-Order Abstract Syntax in
  Coq}}.
\newblock In \bibinfo{editor}{Mariangiola \surnamestart
  Dezani-Ciancaglini\surnameend} \& \bibinfo{editor}{Gordon~D. \surnamestart
  Plotkin\surnameend}, editors: {\sl \bibinfo{booktitle}{TLCA}}, {\sl
  \bibinfo{series}{Lecture Notes in Computer Science}} \bibinfo{volume}{902},
  \bibinfo{publisher}{Springer}, pp. \bibinfo{pages}{124--138}.
\newblock \urlprefix\url{http://dx.doi.org/10.1007/BFb0014049}.

\bibitemdeclare{misc}{fairbairn:alpha}
\bibitem{fairbairn:alpha}
\bibinfo{author}{M. \surnamestart Fairbairn\surnameend}:
  \emph{\bibinfo{title}{A solution to the POPLmark Challenge}}.
\newblock Available at [4].

\bibitemdeclare{article}{DBLP:journals/fac/GabbayP02}
\bibitem{DBLP:journals/fac/GabbayP02}
\bibinfo{author}{M. \surnamestart Gabbay\surnameend} \&
  \bibinfo{author}{A.~M. \surnamestart Pitts\surnameend}
  (\bibinfo{year}{2002}): \emph{\bibinfo{title}{A New Approach to Abstract
  Syntax with Variable Binding}}.
\newblock {\sl \bibinfo{journal}{Formal Asp. Comput.}}
  \bibinfo{volume}{13}(\bibinfo{number}{3-5}), pp. \bibinfo{pages}{341--363}.
\newblock \urlprefix\url{http://dx.doi.org/10.1007/s001650200016}.

\bibitemdeclare{misc}{gacek:abella}
\bibitem{gacek:abella}
\bibinfo{author}{A. \surnamestart Gacek\surnameend}:
  \emph{\bibinfo{title}{A solution to the POPLmark Challenge}}.
\newblock Available at [4].

\bibitemdeclare{phdthesis}{gacek:phd}
\bibitem{gacek:phd}
\bibinfo{author}{A. \surnamestart Gacek\surnameend} (\bibinfo{year}{2009}):
  \emph{\bibinfo{title}{A Framework for Specifying, Prototyping, and Reasoning
  about Computational Systems}}.
\newblock Ph.D. thesis, \bibinfo{school}{University of Minnesota}.

\bibitemdeclare{inproceedings}{DBLP:conf/lics/HarperHP87}
\bibitem{DBLP:conf/lics/HarperHP87}
\bibinfo{author}{R. \surnamestart Harper\surnameend},
  \bibinfo{author}{F. \surnamestart Honsell\surnameend} \&
  \bibinfo{author}{G.~D. \surnamestart Plotkin\surnameend}
  (\bibinfo{year}{1987}): \emph{\bibinfo{title}{A Framework for Defining
  Logics}}.
\newblock In: {\sl \bibinfo{booktitle}{LICS}}, \bibinfo{publisher}{IEEE
  Computer Society}, pp. \bibinfo{pages}{194--204}.

\bibitemdeclare{misc}{hirsch-mag:coq}
\bibitem{hirsch-mag:coq}
\bibinfo{author}{A. \surnamestart Hirschowitz\surnameend} \&
  \bibinfo{author}{M. \surnamestart Maggesi\surnameend}:
  \emph{\bibinfo{title}{A solution to the POPLmark Challenge}}.
\newblock Available at [4].

\bibitemdeclare{article}{DBLP:journals/jar/HirschowitzM12}
\bibitem{DBLP:journals/jar/HirschowitzM12}
\bibinfo{author}{A. \surnamestart Hirschowitz\surnameend} \&
  \bibinfo{author}{M. \surnamestart Maggesi\surnameend}
  (\bibinfo{year}{2012}): \emph{\bibinfo{title}{Nested Abstract Syntax in
  Coq}}.
\newblock {\sl \bibinfo{journal}{J. Autom. Reasoning}}
  \bibinfo{volume}{49}(\bibinfo{number}{3}), pp. \bibinfo{pages}{409--426}.
\newblock \urlprefix\url{http://dx.doi.org/10.1007/s10817-010-9207-9}.

\bibitemdeclare{inproceedings}{Hofmann:1999:SAH:788021.788940}
\bibitem{Hofmann:1999:SAH:788021.788940}
\bibinfo{author}{M. \surnamestart Hofmann\surnameend}
  (\bibinfo{year}{1999}): \emph{\bibinfo{title}{Semantical Analysis of
  Higher-Order Abstract Syntax}}.
\newblock In: {\sl \bibinfo{booktitle}{Proceedings of the 14th Annual IEEE
  Symposium on Logic in Computer Science}}, \bibinfo{series}{LICS '99},
  \bibinfo{publisher}{IEEE Computer Society}, \bibinfo{address}{Washington, DC,
  USA}, pp. \bibinfo{pages}{204--}.
\newblock \urlprefix\url{http://dl.acm.org/citation.cfm?id=788021.788940}.

\bibitemdeclare{inproceedings}{DBLP:conf/icalp/HonsellMS01}
\bibitem{DBLP:conf/icalp/HonsellMS01}
\bibinfo{author}{F. \surnamestart Honsell\surnameend},
  \bibinfo{author}{M. \surnamestart Miculan\surnameend} \&
  \bibinfo{author}{I. \surnamestart Scagnetto\surnameend}
  (\bibinfo{year}{2001}): \emph{\bibinfo{title}{An Axiomatic Approach to
  Metareasoning on Nominal Algebras in HOAS}}.
\newblock In \bibinfo{editor}{Fernando \surnamestart Orejas\surnameend},
  \bibinfo{editor}{Paul~G. \surnamestart Spirakis\surnameend} \&
  \bibinfo{editor}{Jan \surnamestart van Leeuwen\surnameend}, editors: {\sl
  \bibinfo{booktitle}{ICALP}}, {\sl \bibinfo{series}{Lecture Notes in Computer
  Science}} \bibinfo{volume}{2076}, \bibinfo{publisher}{Springer}, pp.
  \bibinfo{pages}{963--978}.
\newblock \urlprefix\url{http://dx.doi.org/10.1007/3-540-48224-5_78}.

\bibitemdeclare{article}{HMS-01}
\bibitem{HMS-01}
\bibinfo{author}{F. \surnamestart Honsell\surnameend},
  \bibinfo{author}{M. \surnamestart Miculan\surnameend} \&
  \bibinfo{author}{I. \surnamestart Scagnetto\surnameend}
  (\bibinfo{year}{2001}): \emph{\bibinfo{title}{pi-calculus in
  (Co)inductive-type theory}}.
\newblock {\sl \bibinfo{journal}{Theor. Comput. Sci.}}
  \bibinfo{volume}{253}(\bibinfo{number}{2}), pp. \bibinfo{pages}{239--285}.
\newblock \urlprefix\url{http://dx.doi.org/10.1016/S0304-3975(00)00095-5}.

\bibitemdeclare{article}{DBLP:journals/entcs/HonsellMS01}
\bibitem{DBLP:journals/entcs/HonsellMS01}
\bibinfo{author}{F. \surnamestart Honsell\surnameend},
  \bibinfo{author}{M. \surnamestart Miculan\surnameend} \&
  \bibinfo{author}{I. \surnamestart Scagnetto\surnameend}
  (\bibinfo{year}{2001}): \emph{\bibinfo{title}{The Theory of Contexts for
  First Order and Higher Order Abstract Syntax}}.
\newblock {\sl \bibinfo{journal}{Electr. Notes Theor. Comput. Sci.}}
  \bibinfo{volume}{62}, pp. \bibinfo{pages}{116--135}.
\newblock \urlprefix\url{http://dx.doi.org/10.1016/S1571-0661(04)00323-8}.

\bibitemdeclare{misc}{leroy:coq}
\bibitem{leroy:coq}
\bibinfo{author}{X. \surnamestart Leroy\surnameend}:
  \emph{\bibinfo{title}{A solution to the POPLmark Challenge}}.
\newblock Available at [4].

\bibitemdeclare{techreport}{Leroy-POPLmark}
\bibitem{Leroy-POPLmark}
\bibinfo{author}{X. \surnamestart Leroy\surnameend} (\bibinfo{year}{2007}):
  \emph{\bibinfo{title}{A locally nameless solution to the {POPLmark}
  challenge}}.
\newblock \bibinfo{type}{Research report} \bibinfo{number}{6098},
  \bibinfo{institution}{INRIA}.

\bibitemdeclare{phdthesis}{mik:eltop}
\bibitem{mik:eltop}
\bibinfo{author}{M. \surnamestart Miculan\surnameend}
  (\bibinfo{year}{1997}): \emph{\bibinfo{title}{Encoding Logical Theories of
  Programs}}.
\newblock Ph.D. thesis, \bibinfo{school}{Dipartimento di Informatica,
  Universit\`a di Pisa, Italy}.

\bibitemdeclare{inproceedings}{DBLP:conf/icfp/MiculanSH05}
\bibitem{DBLP:conf/icfp/MiculanSH05}
\bibinfo{author}{M. \surnamestart Miculan\surnameend},
  \bibinfo{author}{I. \surnamestart Scagnetto\surnameend} \&
  \bibinfo{author}{F. \surnamestart Honsell\surnameend}
  (\bibinfo{year}{2005}): \emph{\bibinfo{title}{Translating specifications from
  nominal logic to CIC with the theory of contexts}}.
\newblock In \bibinfo{editor}{Randy \surnamestart Pollack\surnameend}, editor:
  {\sl \bibinfo{booktitle}{MERLIN}}, \bibinfo{publisher}{ACM}, pp.
  \bibinfo{pages}{41--49}.
\newblock \urlprefix\url{http://doi.acm.org/10.1145/1088454.1088460}.

\bibitemdeclare{article}{DBLP:journals/tocl/MillerT05}
\bibitem{DBLP:journals/tocl/MillerT05}
\bibinfo{author}{D. \surnamestart Miller\surnameend} \&
  \bibinfo{author}{A. \surnamestart Tiu\surnameend} (\bibinfo{year}{2005}):
  \emph{\bibinfo{title}{A proof theory for generic judgments}}.
\newblock {\sl \bibinfo{journal}{ACM Trans. Comput. Log.}}
  \bibinfo{volume}{6}(\bibinfo{number}{4}), pp. \bibinfo{pages}{749--783}.
\newblock \urlprefix\url{http://doi.acm.org/10.1145/1094622.1094628}.

\bibitemdeclare{inproceedings}{DBLP:conf/tlca/Paulin-Mohring93}
\bibitem{DBLP:conf/tlca/Paulin-Mohring93}
\bibinfo{author}{C. \surnamestart Paulin-Mohring\surnameend}
  (\bibinfo{year}{1993}): \emph{\bibinfo{title}{Inductive Definitions in the
  system Coq - Rules and Properties}}.
\newblock In \bibinfo{editor}{Marc \surnamestart Bezem\surnameend} \&
  \bibinfo{editor}{Jan~Friso \surnamestart Groote\surnameend}, editors: {\sl
  \bibinfo{booktitle}{TLCA}}, {\sl \bibinfo{series}{Lecture Notes in Computer
  Science}} \bibinfo{volume}{664}, \bibinfo{publisher}{Springer}, pp.
  \bibinfo{pages}{328--345}.
\newblock \urlprefix\url{http://dx.doi.org/10.1007/BFb0037116}.

\bibitemdeclare{inproceedings}{DBLP:conf/pldi/PfenningE88}
\bibitem{DBLP:conf/pldi/PfenningE88}
\bibinfo{author}{F. \surnamestart Pfenning\surnameend} \&
  \bibinfo{author}{C. \surnamestart Elliott\surnameend}
  (\bibinfo{year}{1988}): \emph{\bibinfo{title}{Higher-Order Abstract Syntax}}.
\newblock In \bibinfo{editor}{Richard~L. \surnamestart Wexelblat\surnameend},
  editor: {\sl \bibinfo{booktitle}{PLDI}}, \bibinfo{publisher}{ACM}, pp.
  \bibinfo{pages}{199--208}.
\newblock \urlprefix\url{http://doi.acm.org/10.1145/53990.54010}.

\bibitemdeclare{book}{DBLP:books/daglib/0005958}
\bibitem{DBLP:books/daglib/0005958}
\bibinfo{author}{B.~C. \surnamestart Pierce\surnameend}
  (\bibinfo{year}{2002}): \emph{\bibinfo{title}{Types and programming
  languages}}.
\newblock \bibinfo{publisher}{MIT Press}.

\bibitemdeclare{article}{DBLP:journals/iandc/Pitts03}
\bibitem{DBLP:journals/iandc/Pitts03}
\bibinfo{author}{A.~M. \surnamestart Pitts\surnameend}
  (\bibinfo{year}{2003}): \emph{\bibinfo{title}{Nominal logic, a first order
  theory of names and binding}}.
\newblock {\sl \bibinfo{journal}{Inf. Comput.}}
  \bibinfo{volume}{186}(\bibinfo{number}{2}), pp. \bibinfo{pages}{165--193}.
\newblock \urlprefix\url{http://dx.doi.org/10.1016/S0890-5401(03)00138-X}.

\bibitemdeclare{misc}{ricciotti:matita}
\bibitem{ricciotti:matita}
\bibinfo{author}{W. \surnamestart Ricciotti\surnameend}:
  \emph{\bibinfo{title}{A solution to the POPLmark Challenge}}.
\newblock Available at [4].

\bibitemdeclare{article}{DBLP:journals/entcs/ScagnettoM02}
\bibitem{DBLP:journals/entcs/ScagnettoM02}
\bibinfo{author}{I. \surnamestart Scagnetto\surnameend} \&
  \bibinfo{author}{M. \surnamestart Miculan\surnameend}
  (\bibinfo{year}{2002}): \emph{\bibinfo{title}{Ambient Calculus and its Logic
  in the Calculus of Inductive Constructions}}.
\newblock {\sl \bibinfo{journal}{Electr. Notes Theor. Comput. Sci.}}
  \bibinfo{volume}{70}(\bibinfo{number}{2}), pp. \bibinfo{pages}{76--95}.
\newblock \urlprefix\url{http://dx.doi.org/10.1016/S1571-0661(04)80507-3}.

\bibitemdeclare{misc}{stump:coq}
\bibitem{stump:coq}
\bibinfo{author}{A. \surnamestart Stump\surnameend}:
  \emph{\bibinfo{title}{A solution to the POPLmark Challenge}}.
\newblock Available at [4].

\bibitemdeclare{manual}{coq}
\bibitem{coq}
\bibinfo{author}{The Coq~Development \surnamestart Team\surnameend}
  (\bibinfo{year}{2011}): \emph{\bibinfo{title}{The Coq Proof Assitant
  Reference Manual, version 8.3}}.
\newblock \bibinfo{organization}{INRIA}.
\newblock \urlprefix\url{http://coq.inria.fr/}.

\bibitemdeclare{misc}{urban:isabelle}
\bibitem{urban:isabelle}
\bibinfo{author}{C. \surnamestart Urban\surnameend\ et al.},
  \emph{\bibinfo{title}{A solution to the POPLmark Challenge}}.
\newblock Available at [4].

\bibitemdeclare{misc}{xi:ats}
\bibitem{xi:ats}
\bibinfo{author}{H. \surnamestart Xi\surnameend}:
  \emph{\bibinfo{title}{A solution to the POPLmark Challenge}}.
\newblock Available at [4].

\end{thebibliography}

\end{document}